\documentclass[aps,reprint,superscriptaddress]{revtex4-1}
\usepackage{CJK}
\usepackage{siunitx}
\usepackage{graphicx}
\usepackage{mathtools,amsthm}
\usepackage{epstopdf}
\usepackage[bookmarks=false,hidelinks]{hyperref}

\begin{document}
\title{Spin-orbit effects at chiral surfaces}
\author{N. K. Lewis}
\affiliation{School of Physics and Astronomy and the Photon Science Institute, University of Manchester, Oxford Road, Manchester, M13 9PL, UK}
\affiliation{The Cockcroft Institute, Daresbury Laboratory, Sci-Tech Daresbury, Warrington, WA4 4AD, UK}
\author{P. J. Durham}
\affiliation{Daresbury Laboratory, Sci-Tech Daresbury, Warrington, WA4 4AD, UK}
\author{W. R. Flavell}
\affiliation{School of Physics and Astronomy and the Photon Science Institute, University of Manchester, Oxford Road, Manchester, M13 9PL, UK}
\author{E. A. Seddon}
\affiliation{School of Physics and Astronomy and the Photon Science Institute, University of Manchester, Oxford Road, Manchester, M13 9PL, UK}
\affiliation{The Cockcroft Institute, Daresbury Laboratory, Sci-Tech Daresbury, Warrington, WA4 4AD, UK}
\date{\today}
\begin{abstract}
Two-dimensional hexagonal and oblique lattices were investigated theoretically with the aim of observing differences in the spin expectation values between chiral and achiral systems. The spin-resolved band structures were derived from the energy eigenvalues and eigenfunctions of a Hamiltonian that includes the lattice potential and the spin-orbit interaction. The spin texture of the achiral hexagonal system was shown to have two non-zero components of the spin polarisation, whereas all three components were calculated to be non-zero for the chiral system. The longitudinal component, found to be zero in the achiral lattice, was observed to invert between the enantiomorphs of the chiral lattice. A heuristic model was introduced to discern the origin of this inverting spin polarisation by considering the dynamics of an electron in chiral and achiral lattices. This model was further used to demonstrate the change in magnitude of the spin polarisation as a function of the lattice parameters and an electric field perpendicular to the lattice.
\end{abstract}
\maketitle
\section{Introduction}
Symmetry-breaking systems have been extensively studied and shown to produce novel spin-polarisation effects \cite{Sunko,Riley}. Experimentally, spin- and angle-resolved photoemission have been used to determine the spin density of two-dimensional systems which lack inversion symmetry \cite{SeddonSpin,Okuda}. Examples of this include the Rashba effect in Au(111) \cite{LaShell}, the giant Rashba effect in Bi/Ag(111)
\cite{Ast,Ast2,Gierz,Sakamoto,Ishizaka} and the spin-valley polarisation present in transition metal dichalcogenide such as WSe$_2$ \cite{Mak,Bertoni,Zhu}. Such phenomena are important for spintronic devices \cite{Dankert,Lo}. However, the spin texture of two-dimensional chiral systems (those that lack mirror symmetry) has had limited study. 

In electron scattering experiments involving gas-phase chiral molecules, spin polarisation inversion has been demonstrated. In particular, Mayer \textit{et al.} showed \cite{Mayer} that longitudinally spin-polarised electrons transmitted through randomly oriented and enantiomerically pure chiral molecules produce an energy-dependent intensity asymmetry between parallel and antiparallel spin polarised electrons. They further observed that the intensity asymmetry `mirrored' between the enantiomers. In another scattering experiment, electrons propagating parallel to the helical axis of DNA have been shown to develop a longitudinal spin polarisation independent of the initial spin orientation \cite{Ray,Gohler}, observations that have been associated with spin filtering effects \cite{Gersten}. These electron scattering experiments (involving systems that lack mirror symmetry) and the electron spin polarisation show a correlation that has been the subject of theoretical investigations \cite{Farago,Cherepkov}.

The main underlying phenomenon that contributes to all these effects is the spin-orbit interaction. Rashba \textit{et al.} considered the spin-orbit interaction in two-dimensional electron gas systems \cite{Manchon}. In the original Rashba model the electronic motion is assumed to be confined to, but free in, the $x$\nobreakdash--$y$~plane and affected by the spin-orbit interaction, where the potential gradient is in the $z$ direction only and approximated to a constant. This model can be solved analytically and two important results arise. Firstly, the two-dimensional energy bands are parabolic but spin-split by an amount proportional to the Rashba parameter, $\alpha$, which is determined by the spin-orbit coupling strength. Secondly, the spin polarisation of each band is entirely in the $x$\nobreakdash--$y$~plane and orthogonal to the surface crystal momentum, $k_{\parallel}$; there is no longitudinal or perpendicular spin polarisation. This model has been used as a phenomenological guide to interpreting experimental results \cite{Gierz} for systems with a relatively free-electron-like band structure. The most rigorous approach currently available is to use fully self-consistent relativistic density-functional methods \cite{Martin}. Such calculations are demanding but can produce good agreement with experiment. However, as Premper \textit{et al.} have suggested \cite{Premper}, useful insight can be gained by a simplified approach. For example, they used a semi-relativistic Hamiltonian with a potential describing the structural symmetry of Bi/Ag(111) to reveal key features of the spin-polarised band structure (which were observed in photoemission experiments \cite{Ast}) and the essence of the effects of the spin-orbit interaction. This is not the only method that can be used to extend the original Rashba model. $\boldsymbol{k}\cdot\boldsymbol{p}$ theory was used to successfully model hexagonal warping effects in Bi$_2$Ti$_3$ \cite{Fu}.

Presented here is a semi-relativistic model that builds upon the work conducted by Premper \textit{et al.} by focusing on two-dimensional lattices with and without mirror symmetry. There are two types of Bravais lattice considered here: the hexagonal lattice, an achiral structure, and the oblique lattice, the only two-dimensional Bravais lattice to lack mirror symmetry. The results show that for the hexagonal lattice there is at least one component of the spin polarisation that is always equal to zero (the longitudinal component) which is true for all achiral lattices. In contrast, all three components are in general non-zero for a chiral lattice. Furthermore, the spin expectation values are found to invert between the enantiomorphs for the longitudinal component, remain unchanged for the tangential component and differ for the out-of-plane component.

\section{Theory}
The band structure and spin texture of two-dimensional Bravais lattices are calculated from the eigenvalues and eigenvectors of the semi-relativistic Hamiltonian (in Hartree units)
\begin{equation}
\hat{H} = \frac{\hat{\boldsymbol{p}}^2}{2m^*} + \hat{V} + \frac{1}{4m^{*2}c^2}\hat{\boldsymbol{\sigma}}\cdot\left(\hat{\boldsymbol{p}}\times\boldsymbol{\nabla}\hat{V}\right),
\label{E2}
\end{equation}
where $m^*$ is the effective electron mass, $c$ is the speed of light in a vacuum, $\hat{\boldsymbol{p}}$ is the momentum operator, $\hat{\boldsymbol{\sigma}}$ is the vector of Pauli matrices and $\hat{V}$ is the potential operator \cite{Hermitian}. Although the electron motion is confined to the $x$\nobreakdash--$y$~plane, the potential gradient, $\boldsymbol{\nabla}V$, spans three-dimensional space with $x$, $y$ and $z$ components. The potential is a periodic function in the $x$\nobreakdash--$y$~plane described by a Fourier series given by
\begin{equation}
V(\boldsymbol{r}) = \sum_{\boldsymbol{G}}V_{\boldsymbol{G}}e^{i\boldsymbol{G}\cdot\boldsymbol{r}}, \quad \boldsymbol{r} = (x,y)
\label{E1}
\end{equation}
where $\boldsymbol{G}$ is a two-dimensional reciprocal lattice vector and the $V_{\boldsymbol{G}}$ are in general a set of complex variables whose properties are described in the supplemental information \cite{Supp}. This calculation goes beyond the standard Rashba model by including the potential and its in-plane gradients in the $x$ and $y$ directions which are derived analytically from Eq. \ref{E1}. The gradient normal to the plane (\textit{i.e.} in the $z$ direction) is assumed to be a constant given by the standard Rashba parameter:
\begin{equation}
\alpha = \frac{1}{4m^{*2}c^2}\frac{\partial V}{\partial z}.
\end{equation}
Thus the reciprocal lattice vectors, the corresponding Fourier components $V_{\boldsymbol{G}}$ and the Rashba parameter $\alpha$ are the principal input parameters to the calculations. 

Given a Hamiltonian in the form of Eq. \ref{E2} and a periodic potential, it is natural to use a basis formed from the tensor product of momentum eigenkets, $|\boldsymbol{k}\rangle$, and spin $\hat{S}_z$ eigenkets, $|m_s\rangle$. In this basis, the Bloch ket for a given $\boldsymbol{k}$ can be written as
\begin{equation}
|\Psi(\boldsymbol{k})\rangle = \sum_{\boldsymbol{g},m_s}u_{\boldsymbol{g},m_s}(\boldsymbol{k})|\boldsymbol{k}+\boldsymbol{g},m_s\rangle,
\end{equation} 
where $\boldsymbol{g}$ is a reciprocal lattice vector. The eigenvectors $u_{\boldsymbol{g},m_s}(\boldsymbol{k})$ and the eigenvalues $E(\boldsymbol{k})$ are found by diagonalising the Hamiltonian matrix in the $|\boldsymbol{k}+\boldsymbol{g},m_s\rangle$ basis, using the standard techniques of linear algebra \cite{Press}. The spin expectation values $\langle S_{x,y,z}\rangle$ are calculated by constructing the spin density matrix from the full eigenvector $u_{\boldsymbol{g},m_s}(\boldsymbol{k})$ for a given $\boldsymbol{k}$.

This generalised Rashba calculation is a tool with which the parameter space of the model can be easily scanned. The main focus of the work reported here is to observe the effects of chirality on spin textures, an objective that requires a lattice potential to be included in the model, hence demanding a two-dimensional lattice structure to be specified. Any periodic system in the $x$\nobreakdash--$y$~plane can be constructed from one of the five two-dimensional Bravais lattices. The basis placed on each lattice site may be simply a single atom or a multi-atom cluster or molecule (which may itself be chiral). Of these Bravais lattices, only the so-called oblique lattice has no ($x\to -x)$ or $(y\to-y)$ reflection symmetries; all the others (rectangular, centred rectangular, square and hexagonal) possess these symmetries. Thus the oblique lattice is called a chiral system while the other four are achiral. In this paper, calculations were performed on the oblique lattice itself (not a multi-atom unit cell) as a function of its structural parameters without varying the potential parameters. This is because special values of these potential parameters correspond to a particular hexagonal lattice (see the following section). Varying the structural parameters continuously provides a convenient way of transforming an achiral system into a chiral system. In this way, the effects of chirality on the spin textures can be observed and investigated.

\section{Computational details}
The input parameters used in the calculations reported below were based on the surface of Bi/Ag(111) described in reference \cite{Premper}. The Fourier components of the potential were assigned the values $V_{0} = -6.6\,\,\si{eV}$ and $V_{\boldsymbol{G}} = \pm i5\,\,\si{eV}$ which associate with $\boldsymbol{G} = 0$ and the shortest reciprocal lattice vectors, respectively. Except for $V_{\boldsymbol{G}=0}$, adjacent Fourier components were always of opposite sign such that all lattices had an antisymmetric potential that breaks local inversion symmetry \cite{Premper}. This asymmetry is a consequence of the \textit{ABC} stacking present in face-centered cubic (111) surfaces which was retained through the achiral lattices modelled. The other parameters used were an effective electron mass of $-0.4m_e$, a Rashba coefficient of $0.392\,\,\si{eV\,\,\angstrom}$ and a scaled in-plane potential gradient \cite{Premper}. Tests confirming that the model correctly reproduces the work conducted by Premper \textit{et al}. are shown in the supplemental information \cite{Supp}. 

Figure \ref{F1} shows the direct lattice points of an oblique structure from which all reciprocal lattices and the corresponding reciprocal lattice vectors, $\boldsymbol{g}$, were derived.
\begin{figure}
 	\includegraphics[width=\linewidth]{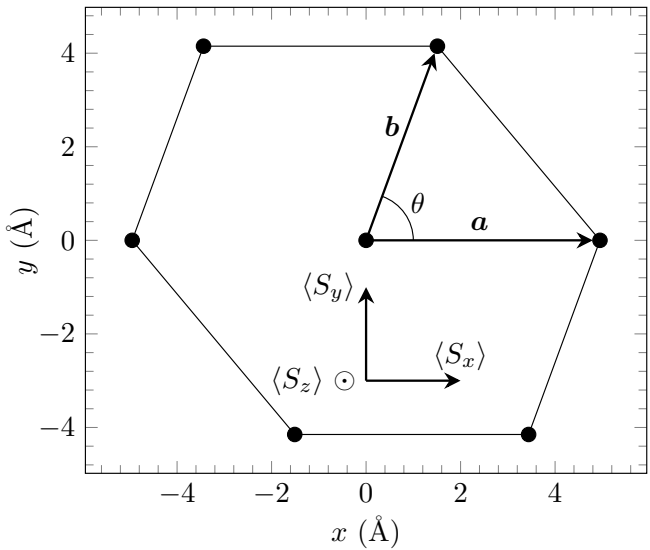}
 	\caption{Direct lattice points for an oblique lattice, where $\boldsymbol{a}$ and $\boldsymbol{b}$ are the primitive lattice vectors. $|\boldsymbol{a}|$ was set constant to $4.95\,\,\si{\angstrom}$ for all calculations. The directions of the spin expectation values are shown relative to the real-space directions.}
 	\label{F1}
\end{figure}
For arbitrary $\theta$ and $|\boldsymbol{a}|/|\boldsymbol{b}|$ the chirality of this structure is evident. The first calculation presented used the hexagonal structure of Bi/Ag(111) produced by setting $|\boldsymbol{a}| = |\boldsymbol{b}| = 4.95\,\,\si{\angstrom}$ and $\theta = 60\,\si{\degree}$ where $|\boldsymbol{b}|$ and $\theta$ are called the structural parameters. The results of this calculation were compared with that of a chiral lattice with the structural parameters $|\boldsymbol{b}| = |\boldsymbol{a}| - 1\,\,\si{\angstrom}$ and $\theta = 60\,\si{\degree}$. To then observe variations in the spin texture, different chiral lattices were used. These were obtained by varying the structural parameters: $|\boldsymbol{b}|$ was changed for constant $\theta = 60\,\si{\degree}$ and $\theta$ was varied for constant $|\boldsymbol{b}| = |\boldsymbol{a}|-1\,\,\si{\angstrom}$. The complementary enantiomorph of a particular chiral lattice was generated by inverting the $\hat{\boldsymbol{x}}$ component for all lattice vectors.

The eigenvectors and eigenvalues reported in the next section were calculated as a function of $k_x$ for the states $|\Psi(k_x,k_y = 0)\rangle$. In this case, the spin polarisations are labelled as longitudinal for $\langle S_x\rangle$, tangential for $\langle S_y\rangle$ and perpendicular for $\langle S_z\rangle$. The directions of the spin are indicated in Fig. \ref{F1}. Calculations were also performed as a function of $k_y$ for the states $|\Psi(k_x = 0,k_y)\rangle$ and the results of these are included where appropriate. For this state, the longitudinal spin direction is $\langle S_y\rangle$ and the tangential direction is $\langle S_x\rangle$.
\section{Results}
\begin{figure}
	\includegraphics[width=\linewidth]{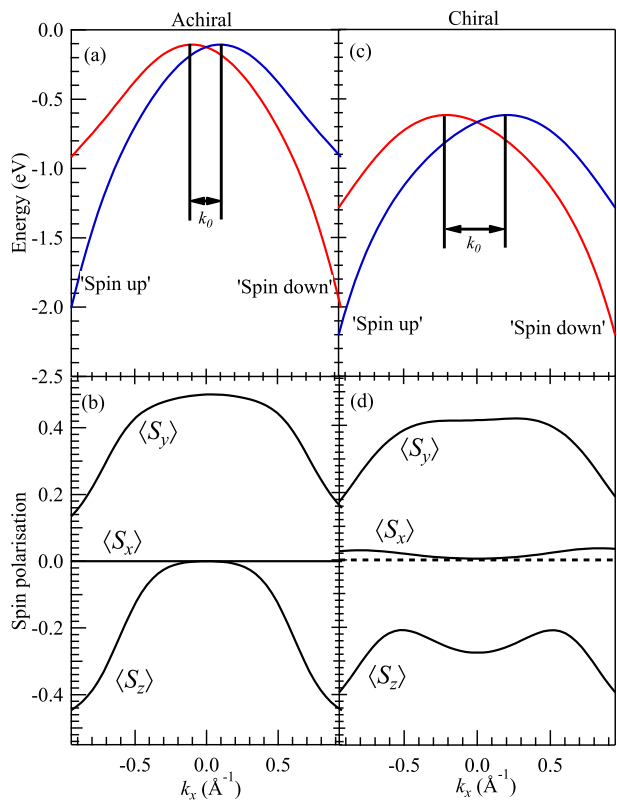}
	\caption{(a) the lowest-energy eigenvalues for the states $|\Psi(k_x,k_y = 0)\rangle$ of a hexagonal lattice where $k_0$ describes the Rashba splitting. (c) shows the same eigenvalues for the oblique lattice. The first Brillouin zone boundaries in the $k_x$ direction for the hexagonal and oblique lattices are at $\pm0.88\,\,\si{\angstrom^{-1}}$ and $\pm0.87\,\,\si{\angstrom^{-1}}$, respectively (see supplemental information for an extended energy eigenvalue diagram). The spin expectation values of the `spin up' band in (a) and (c) are shown in (b) and (d), respectively. The spin expectation values of the `spin down' were calculated and agree with Kramer's theorem \cite{Ashcroft} (not shown).}
	\label{F2}
\end{figure}

Figures \ref{F2}(a) and (b) show the lowest energy eigenvalues and the spin expectation values of the `spin up' band, for $|\Psi(k_x,k_y = 0)\rangle$ of the achiral hexagonal lattice. The bands are labelled as `spin up' and `spin down' because the signs of $\langle S_y\rangle$ are positive and negative respectively. The main difference between the values shown in Fig. \ref{F2}(b) and those of the original formulation of the Rashba effect is that $\langle S_z\rangle$ is not zero for all $k_x$. This is caused by the non-zero in-plane potential gradients. The longitudinal spin polarisation, $\langle S_x\rangle$, is observed to be zero for all $k_x$. This can be rationalised by considering the hexagonal lattice to be an equal mixture of both enantiomorphs. The size of the band splitting, $k_0$, in Fig. \ref{F2}(a) agrees with that reported previously \cite{Ast,Premper}.

The lowest energy eigenvalues for $|\Psi(k_x,k_y = 0)\rangle$ of the chiral oblique lattice are shown in Fig. \ref{F2}(c). The spin splitting is approximately twice as large as that in Fig. \ref{F2}(a). Even though the chiral system lacks mirror symmetry, the bands show a symmetry about $k_x = 0\,\,\si{\angstrom^{-1}}$. This is because a rotation of $180\,\si{\degree}$ about the reciprocal lattice point at $(0,0)$ generates the equivalent chiral lattice.   

The spin expectation values of the `spin up' band in Fig. \ref{F2}(c) are shown in Fig. \ref{F2}(d). There are two important differences in the spin expectation values between the achiral (Fig. \ref{F2}(b)) and chiral lattices (Fig. \ref{F2}(d)). Firstly, there is an increase in the magnitude of $\langle S_z\rangle$ in the region $-0.5<k_x<0.5\,\,\si{\angstrom^{-1}}$. This is a result of the in-plane components of the potential gradient and the chirality of the lattice. Secondly, there is a non-zero longitudinal spin polarisation shown by $\langle S_x\rangle$ for the chiral lattice. It is positive, small in magnitude ($\sim 2.5\%$ at its maximum) and varies as a function of $k$. This result is due to the oblique lattice lacking mirror symmetry and the perpendicular component of the potential gradient.

\begin{figure}
	\centering
	\includegraphics[width=\linewidth]{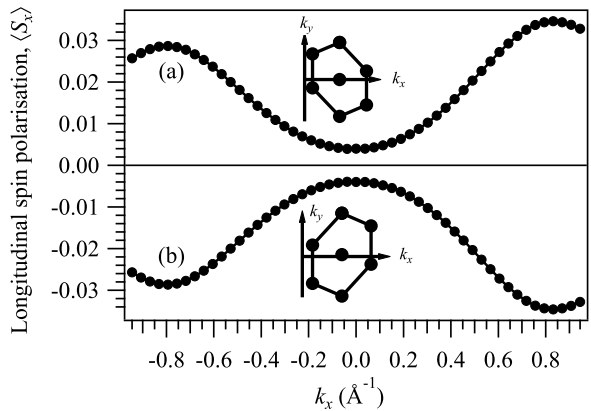}
	\caption{Longitudinal spin polarisation, $\langle S_x\rangle$, obtained from the `spin up' band of the lowest energy eigenvalues for the states $|\Psi(k_x,k_y = 0)\rangle$; (a) a magnified image of $\langle S_x\rangle$ from Fig. \ref{F2}(d) calculated using the displayed upper-reciprocal lattice, and (b) $\langle S_x\rangle$ for the mirror-reflected structure shown.}  
	\label{F3}
\end{figure}
Figure \ref{F3} shows the longitudinal spin polarisation as a function of $k_x$, where (a) shows $\langle S_x\rangle$ from Fig. \ref{F2}(d) but with an increased resolution and (b) shows $\langle S_x\rangle$ for the `spin up' band of the complementary chiral lattice (the mirror-reflected lattice). Clearly, the longitudinal spin polarisation is equal in magnitude but opposite in sign for the enantiomorphs. The other spin polarisations, $\langle S_y\rangle$ and $\langle S_z\rangle$, remain unchanged between the enantiomorphs. Spin expectation values were also calculated for the states $|\Psi(k_x = 0, k_y)\rangle$ (\textit{i.e.} with $\boldsymbol{k}$ along the $y$ direction), and these show that $\langle S_y\rangle$ inverts between the enantiomorphs whereas $\langle S_x\rangle$ does not. Furthermore, in contrast to the calculations associated with Fig. \ref{F3}, $\langle S_z\rangle$ is also found to invert between the enantiomorphs.

The influence of chirality on the spin polarisations is further revealed by varying the structural parameters and the magnitude of the perpendicular component of the potential gradient, $\partial_zV = E_z$. An average longitudinal spin polarisation, $\langle \bar{S}_x\rangle$, for the `spin up' band (of the same enantiomorph) was calculated over the range $k_y = 0$ to the first Brillouin zone boundary for $|\Psi(k_x=0, k_y)\rangle$. This was performed using oblique lattices with varying values of the structural parameters $|\boldsymbol{b}|$ and $\theta$ (see section III)

\begin{figure}
	\centering
	\includegraphics[width=\linewidth]{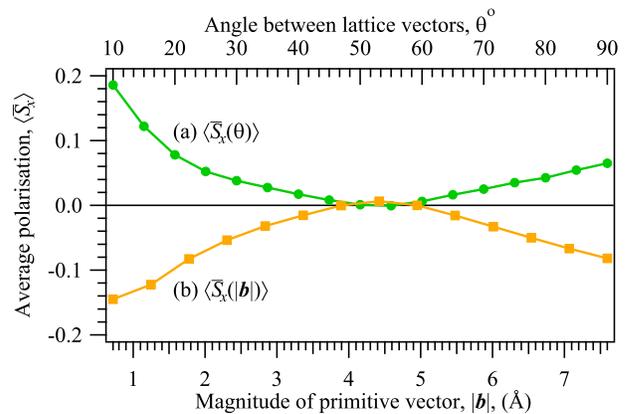}
	\caption{The average longitudinal spin polarisation as a function of (a) the angle between the primitive lattice vectors, and (b) the magnitude of the primitive lattice vector $|\boldsymbol{b}|$.}
	\label{F4}
\end{figure}

In figure \ref{F4}, (a) shows the variation of the average longitudinal spin polarisation against increasing $\theta$, and (b) shows the variation with the magnitude of the primitive-lattice vector, $|\boldsymbol{b}|$. 
Note that for $|\boldsymbol{b}| = |\boldsymbol{a}| = 4.95\,\,\si{\angstrom}$ and $\theta = 60\,\si{\degree}$ the oblique lattice becomes hexagonal. Figures \ref{F4}(a) and (b) thus show that the average longitudinal spin polarisation increases as the values of either of the structural parameters diverge from their respective hexagonal structure values. The same calculation was performed for the complementary enantiomorph and produced the same variation but inverted (not shown).

\begin{figure}
	\centering
	\includegraphics[width=\linewidth]{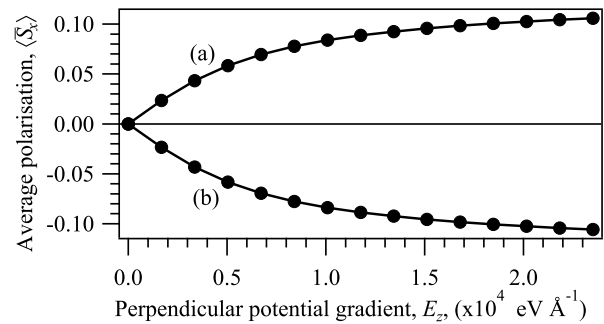}
	\caption{Average longitudinal spin polarisation as a function of the magnitude of the perpendicular component of the potential gradient. The average polarisations were calculated using the same procedure as in Fig. \ref{F4}. (a) labels the dependence for the enantiomorph associated with Fig. \ref{F4}, and (b) corresponds to the mirror-reflected lattice.}
	\label{F5}
\end{figure}

The variation of the average longitudinal spin polarisation as a function of $E_z$ is displayed in Fig. \ref{F5}.
This shows that for small $E_z$ the longitudinal spin polarisation is proportional to the perpendicular potential gradient. Figure 5 also infers the relationship between the spin-orbit coupling and the longitudinal spin polarisation. This is because the spin-orbit coupling is derived using the perpendicular potential gradient such that for a spherical potential $E_z \propto Z/r^2$. Hence, increasing $E_z$ is equivalent to increasing the spin-orbit coupling (and the atomic number, $Z$,) which corresponds to an increased longitudinal spin polarisation. In the next section, a heuristic model is put forward to understand this effect.

\section{Discussion}
In the original Rashba model, where the in-plane electronic motion is approximated to be free and the only potential gradient is perpendicular to the plane, the spin-orbit interaction from Eq. \ref{E2} becomes $\alpha\boldsymbol{\sigma}\cdot\left(\boldsymbol{k}_{\parallel}\times\hat{\boldsymbol{z}}\right)$. Thus the spin polarisation is clearly tangential to the surface crystal momentum, $k_{\parallel}$, and there is no longitudinal or perpendicular spin polarisation. In contrast, when an in-plane potential is included, the calculations show non-zero longitudinal, $\langle S_x\rangle$, and perpendicular, $\langle S_z\rangle$, polarisations which are shown in Figs. \ref{F2}(b) and (d). Finite values of $\langle S_z\rangle$ arise because of the non-zero in-plane potential gradients and antisymmetric Fourier coefficients. However, the appearance of a longitudinal spin polarisation is less intuitive. This longitudinal polarisation can be understood in terms of the following heuristic model.

\begin{figure}
	\includegraphics[width=\linewidth]{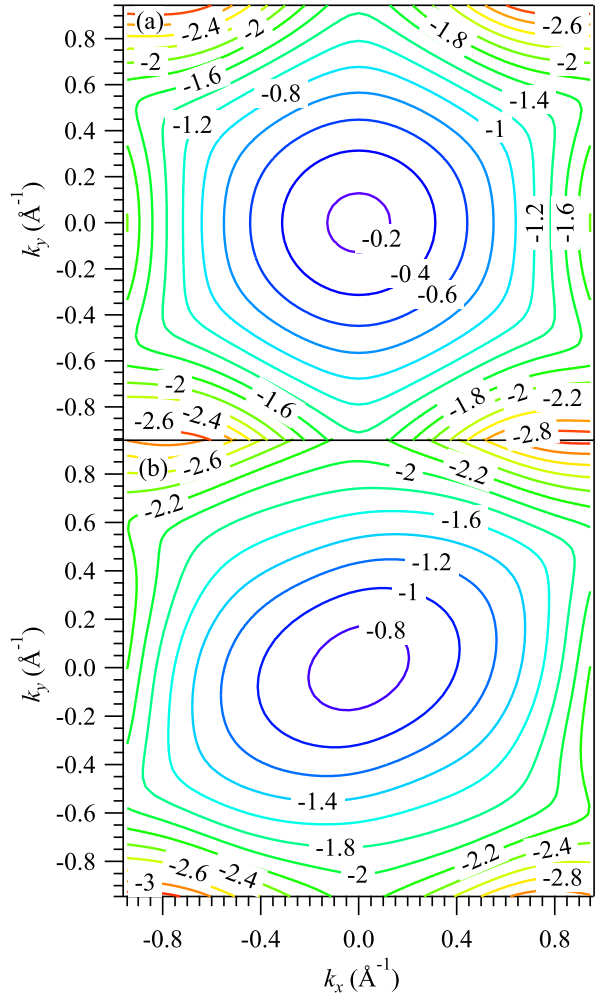}
	\caption{Constant energy eigenvalues (in eV) as a function of $k_x$ and $k_y$ for (a) the hexagonal lattice, and (b) one enantiomorph of the oblique lattice. These were obtained from the energy eigenvalues shown in Fig. \ref{F2}.}
	\label{F7}
\end{figure}

Consider the effective velocities for Bloch electrons in lattices of different symmetries. The traditional approach to electron dynamics \cite{Ashcroft,Ziman} associates this velocity, $\boldsymbol{v}$, with the momentum derivative of the energy eigenfunctions, $E(\boldsymbol{k})$,
\begin{equation}
\boldsymbol{v}(\boldsymbol{k}) = \frac{\partial E(\boldsymbol{k})}{\partial \boldsymbol{k}}.
\label{E5}
\end{equation}
The energy contours shown in Fig. \ref{F7}(a) indicate that for the hexagonal lattice this derivative is zero for $k_x = 0$ and $k_y = 0$ for all bands.
However, this is not the case for the chiral lattice as shown in Fig. \ref{F7}(b). Furthermore, the momentum-space gradient of the energy eigenvalues in Eq. \ref{E5} are equal to momentum expectation values in the corresponding Bloch state \cite{Ashcroft}:
\begin{equation}
\nabla_{\boldsymbol{k}} E(\boldsymbol{k}) = \left(m^*\right)^{-1}\langle\Psi(\boldsymbol{k})|\hat{\boldsymbol{p}}|\Psi(\boldsymbol{k})\rangle.
\end{equation}
Thus the calculated energy contours in Fig. \ref{F7}(a) imply that a Bloch state $|\Psi(k_x,k_y = 0)\rangle$ is associated with a velocity $\boldsymbol{v}(k_x,k_y = 0)$ which is entirely along the $x$ direction for the hexagonal lattice, but has a component along the $y$ axis for the chiral lattice. Moreover, it can be shown analytically that for a lattice with mirror symmetry the momentum expectation value $\langle \Psi(k_x,k_y=0)|\hat{p}_y|\Psi(k_x,k_y=0)\rangle$ must always be zero. Hence, our calculations confirm the general results: for the chiral lattice
\begin{align}
\left(m^*\right)^{-1}\langle\Psi(k_x,&k_y = 0)|\hat{p}_y|\Psi(k_x,k_y = 0)\rangle \nonumber \\ &= v_y(k_x,k_y = 0)\neq 0, \label{E9b}
\end{align}
and for the achiral lattice
\begin{align}
\left(m^*\right)^{-1}\langle\Psi(k_x,&k_y = 0)|\hat{p}_y|\Psi(k_x,k_y = 0)\rangle \nonumber \\ &= v_y(k_x,k_y = 0) = 0. \label{E9a}
\end{align}
This velocity is now treated as a classical variable and a Lorentz transformation is performed from the lab frame to the rest frame of an electron in the Bloch state $|\Psi(k_x,k_y = 0)\rangle$. This electron is taken to have a velocity $\boldsymbol{v} = c(\beta_x,\beta_y)$, where $\beta_y$ is non-zero only in a chiral system (and $c$ is the speed of light). In the lab frame, with no magnetic field present, there exists a static electric field, $\boldsymbol{E}$, given by the potential gradient. In the electron's rest frame a magnetic field $\boldsymbol{B}'$ appears \cite{Jackson} with the in-plane components
\begin{equation}
B_y' = \gamma\beta_x\frac{E_z}{c} \quad \text{and} \quad B_x' = \pm\gamma\beta_y\frac{E_z}{c},
\label{E8}
\end{equation} 
where $\gamma = (1+\beta^2)^{-1/2}$ is the Lorentz factor. The electron spin will align with this magnetic field; this is the ``semi-classical'' picture of the origin of the spin-orbit interaction. The $B'_y$ component shown in Eq. \ref{E8} is similar to the spin polarisation calculated from the Rashba model; it is tangential to the non-zero momentum eigenket $|k_x,k_y = 0\rangle$ and the perpendicular electric field. There is also a magnetic field component along the $x$ direction in Eq. \ref{E8}, hence a finite longitudinal spin polarisation is expected. The heuristic model proposed suggests that the longitudinal polarisation shown in Fig. \ref{F2}(d) is due to the chiral lattice lacking mirror symmetry which then results in a non-zero $\beta_y$ velocity determined by Eq. \ref{E9b}. The velocity of the electron in the $y$ direction, $\beta_y$, can be positive or negative such that $B'_x$ displayed in Eq. $\ref{E8}$ has been modified with a pre-factor of $\pm$ representing the handedness of the enantiomorphs. Figure \ref{F3} shows that the longitudinal spin polarisation does invert between the two enantiomorphs.

Equation \ref{E8} also suggests that the longitudinal spin polarisation should be proportional to both the velocity $\beta_y$, and the perpendicular electric field $E_z$ for small $B'_x$. The velocity $\beta_y$ depends on the structural parameters $\theta$ and $|\boldsymbol{b}|$ shown in Fig. \ref{F1} because these parameters allow the lattice to transform from achiral to chiral. In figure \ref{F4}, (a) shows that at $\theta = 60\,\si{\degree}$ the spin polarisation is nominally zero because the lattice points are approximately hexagonal as $|\boldsymbol{b}|= |\boldsymbol{a}| - 1\,\,\si{\angstrom}$. Therefore, as $\theta$ departs from $60\,\si{\degree}$ the longitudinal spin polarisation increases because the structure is distorted further from the hexagonal symmetry. Similarly, the same relationship is observed when the magnitude of the basis vector, $|\boldsymbol{b}|$, is varied as shown by curve (b) in Fig. \ref{F4}. At $|\boldsymbol{b}| = 4.95\,\,\si{\angstrom}$ the lattice is hexagonal and the longitudinal spin polarisation is zero. As $|\boldsymbol{b}|$ is varied from $4.95\,\,\si{\angstrom}$ the longitudinal spin polarisation increases. Equation \ref{E8} shows that the longitudinal spin polarisation is zero for $E_z = 0$ (as $B_x' = 0$) and is expected to evolve linearly as $E_z$ increases. These effects are both observed from the initial three points shown in Fig. \ref{F5}.

When the structural distortion of the chiral lattice from hexagonal becomes large, the longitudinal spin polarisation is no longer a linear function of $\beta_y$. Similarly, when the perpendicular electric field, $E_z$, becomes large, deviations are found (see Fig. \ref{F5}) from the linear relationship in Eq. \ref{E8}. Such detailed behaviour would presumably be reproduced by modern \textit{ab initio} methods such as relativistic-DFT calculations, but are beyond this heuristic model. 

The perpendicular spin expectation values $\langle S_z\rangle$ shown in Figs. \ref{F2}(b) and (d) are also explained by the heuristic model introduced above. Using the same derivation that produced the in-plane magnetic field components displayed in Eq. \ref{E8}, the perpendicular component is \cite{Jackson}
\begin{equation}
B_z' = -\frac{\gamma}{c}\left(\beta_xE_y - \beta_yE_x\right),
\label{E7}
\end{equation} 
where $E_x$ and $E_y$ are the in-plane electric fields. Furthermore, for an electron associated with $|\Psi(k_x, k_y = 0)\rangle$, the chiral contribution comes from the term $\beta_yE_x$, since $\beta_y \neq 0$ for the chiral lattice. Similarly, for $|\Psi(k_x = 0, k_y)\rangle$ the chiral contribution is from $\beta_xE_y$. This implies that $\langle S_z\rangle$ for the states $|\Psi(k_x,k_y = 0)\rangle$ of the chiral lattice has no chiral contribution because $E_x$ is zero (see the supplemental information \cite{Supp}). Therefore, $\langle S_z\rangle$ does not change between the two enantiomorphs as obtained from the calculations. However, for $|\Psi(k_x = 0,k_y)\rangle$ the values of $\langle S_z\rangle$ were found to invert between the two enantiomorphs. This is because the only non-negligible term \cite{Supp} contributing to the perpendicular magnetic field comes from $\beta_xE_y$ which is the chiral factor. For a more general chiral lattice $\langle S_z\rangle$ is different between the enantiomorphs as both terms in Eq. \ref{E7} will be non-zero.

The non-zero perpendicular spin polarisation shown in Fig. \ref{F2}(d) at $k_x = 0$ which is zero in Fig. \ref{F2}(b) is another consequence of the lack of mirror symmetry of the oblique lattice. To see this, Eq. \ref{E9a} is rewritten using $\hat{p}_x$ to produce the equivalent equality for the achiral lattice
\begin{equation}
\langle\Psi(k_x = 0,k_y|\hat{p}_x|\Psi(k_x = 0,k_y)\rangle = 0,
\end{equation}
which is true for any $k_y$ including $k_y = 0$. This implies that $\beta_x = 0$ at $k_x = 0$ for an achiral lattice but not in general for the chiral lattice. The maximum in absolute value of $\langle S_z\rangle$ shown in Fig. \ref{F2}(d) at $k_x = 0$ is a result of the in-plane electric field $E_y$ (the functional form of which is derived in the supplemental information \cite{Supp}). Therefore, at $k_x = k_y = 0$ the magnitude of the electric field is maximised resulting in a large perpendicular spin polarisation component.

Figures \ref{F2}(b) and (d) show that $\langle S_y\rangle \neq 0$ at $k_x = k_y = 0$ even though Eq. \ref{E8} shows that $B_y \propto \beta_x$. This can be explained by noticing that the eigenfunctions derived from either the simple Rashba model or the full calculations, which include an in-plane potential, contain a phase that is absent from the semi-classical heuristic model described in this section. This phase produces the tangential spin components \cite{Manchon}. 

Experimental observation of the longitudinal spin polarisation should be possible using spin-resolved photoemission \cite{Park,Sanchez,Pan}, as several different types of chiral surfaces exist \cite{Jenkins}. Such experiments are to be distinguished from spin-resolved photoemission experiments where the chirality is a result of the experimental geometry, as explored in Ref. \cite{Kobayashi}. The key role of the spin-orbit interaction implies that surfaces composed of heavy atoms are expected to produce a significant longitudinal spin polarisation. To determine whether the longitudinal spin polarisation is energetically resolvable in spin-resolved photoemission experiments, the above heuristic model can be used to calculate an energy shift, $\Delta E$, associated to the longitudinal polarisation and its corresponding magnetic field component, $B_{\text{long}}$. This energy shift is obtained from  $\Delta E = g_s\mu_BB_{\text{long}}$, where $\mu_B$ is the Bohr magneton and $g_s$ is the electron g-factor. Using the Rashba parameter appropriate for Bi/Ag(111) and a momentum expectation value associated with the longitudinal spin polarisation, an energy shift of $24\,\,\si{meV}$ is obtained. Below $100\,\,\si{K}$ thermal fluctuations are an insignificant factor in randomising electron spins with respect to the longitudinal direction and as the total instrumental resolution of state-of-the-art spin-resolved photoemission experiments at synchrotrons is $\sim 70\,\,\si{meV}$ \cite{Chiara} or better \cite{Okuda11} this will allow for the polarisation of the surface states to be resolved in the future.

\section{Conclusion}
The Rashba model has been generalised, extending the approach of Premper \textit{et al.}, to include lattices of different symmetries, the lattice potential and its in-plane gradients. This was done using a basis of spin and momentum eigenkets, and a potential described by a set of Fourier components. The in-plane potential gradients included in the spin-orbit interaction were derived from these parameters. The perpendicular component was approximated to a constant as performed in the Rashba model. This parametrised model was solved numerically by standard linear algebra techniques to derive the energy eigenfunctions, eigenvectors and the spin expectation values. 

The main focus of this paper has been to use this parametrised model to investigate the effects of lattice symmetry on the spin-dependent electronic structure of the system. A simple two-dimensional chiral lattice was chosen for study, the oblique Bravais lattice, which has no mirror symmetries except for special values of its basis vectors when it becomes, for example, the hexagonal Bravais lattice. The calculations show that chirality is associated with the appearance of a longitudinal spin polarisation and modifications of the perpendicular polarisation. 

To provide physical insight into these findings, a heuristic model has been suggested utilising a semi-classical approach to electron dynamics: a velocity is associated with a given Bloch state and is then used in a Lorentz transformation. This heuristic model explains key differences between the behaviour of chiral and achiral systems, and serves as a guide to observations, experimental and theoretical, on more complex systems.
\begin{acknowledgments}
	This work was supported by EPSRC (UK) under Grant number EP/M507969/1. Funding was also received from ASTeC and the Cockcroft Institute (UK) The data associated with the paper are openly available from Mendeley: \url{http://dx.doi.org/10.17632/sjvrp2ft2x.1}.
\end{acknowledgments}


\begin{thebibliography}{38}
\bibitem [{Her()}]{Hermitian}%
	\BibitemOpen
	\href@noop {} {}\bibinfo {note} {{This form emits the Darwin and
			mass-velocity terms, hence in general it is not Hermitian. This is accounted
			for in the numerical calculations.}}\BibitemShut {Stop}%
\end{thebibliography}
\end{document}